\font\sixrm=cmr6
\font\sixi=cmmi6
\font\sixsy=cmsy6

\font\sevenrm=cmr7
\font\seveni=cmmi7
\font\sevensy=cmsy7

\font\tenrm=cmr10
\font\teni=cmmi10
\font\tensy=cmsy10
\font\tenit=cmti10
\font\tensl=cmsl10
\font\tenbf=cmbx10
\font\tentt=cmtt10

\font\twelverm=cmr12
\font\twelvei=cmmi12
\font\twelvesy=cmsy10 at 12pt
\font\twelveit=cmti12
\font\twelvesl=cmsl12
\font\twelvebf=cmbx12
\font\twelvett=cmtt12

\def\twelvepoint{%
\def\rm{\fam0\twelverm}%
\def\it{\fam\itfam\twelveit}%
\def\sl{\fam\slfam\twelvesl}%
\def\bf{\fam\bffam\twelvebf}%
\def\tt{\fam\ttfam\twelvett}%
\def\cal{\twelvesy}%
 \textfont0=\twelverm
  \scriptfont0=\sevenrm
  \scriptscriptfont0=\sixrm
 \textfont1=\twelvei
  \scriptfont1=\seveni
  \scriptscriptfont1=\sixi
 \textfont2=\twelvesy
  \scriptfont2=\sevensy
  \scriptscriptfont2=\sixsy
 \textfont3=\tenex
  \scriptfont3=\tenex
  \scriptscriptfont3=\tenex
 \textfont\itfam=\twelveit
 \textfont\slfam=\twelvesl
 \textfont\bffam=\twelvebf
 \textfont\ttfam=\twelvett
 \baselineskip=15pt
}
\def\tenpoint{%
\def\rm{\fam0\tenrm}%
\def\it{\fam\itfam\tenit}%
\def\sl{\fam\slfam\tensl}%
\def\bf{\fam\bffam\tenbf}%
\def\tt{\fam\ttfam\tentt}%
\def\cal{\tensy}%
 \textfont0=\tenrm
  \scriptfont0=\sevenrm
  \scriptscriptfont0=\sixrm
 \textfont1=\teni
  \scriptfont1=\seveni
  \scriptscriptfont1=\sixi
 \textfont2=\tensy
  \scriptfont2=\sevensy
  \scriptscriptfont2=\sixsy
 \textfont3=\tenex
  \scriptfont3=\tenex
  \scriptscriptfont3=\tenex
 \textfont\itfam=\tenit
 \textfont\slfam=\tensl
 \textfont\bffam=\tenbf
 \textfont\ttfam=\tentt
 \baselineskip=12pt
}

\font\authorrm=cmr12 scaled \magstep1

\font\titlebx=cmbx12 scaled \magstep2


\hsize     = 148mm
\vsize     = 236mm
\hoffset   =    5mm
\voffset   =    4mm
\topskip   =  19pt
\parskip   =   0pt
\parindent =   0pt

\newskip\one
\one=15pt

\newcount\LastMac
\def\Skipe{1}  

\def\SkipToFirstLine{
 \LastMac=\Skipe
 \dimen255=150pt
 \advance\dimen255 by -\pagetotal
 \vskip\dimen255
}

\def\Raggedright{%
 \rightskip=0pt plus \hsize
 \spaceskip=.3333em
 \xspaceskip=.5em
}

\def\Fullout{
 \rightskip=0pt
 \spaceskip=0pt
 \xspaceskip=0pt
}


\def\ct#1\par{
 \titlebx\baselineskip=22pt
 #1
\vskip0.8truecm
}

\def\ca#1\par{
 \authorrm\baselineskip=18pt
 #1
\vskip0.8truecm
}

\def\aa#1\par{
 \twelveit\baselineskip=15pt
 #1
\vskip0.5truecm
}

\def\ha#1\par{
 \Raggedright
\vskip15pt
 \twelvebf\baselineskip=15pt
 #1

\noindent
}

\def\hb#1\par{
\vskip15pt
 \twelveit\baselineskip=15pt
 #1

\noindent
}

\def\tx{
 \Fullout
 \twelvepoint\rm
}
\def\tf{
 \Fullout
 \lineskip=12pt
 \tenpoint\rm
}


\nopagenumbers

\input psfig.tex
\headline={\twelverm 16th ECRS OG-4.5  \hfil  Page \folio}
\def\part#1#2{{{\partial {#1}}\over {\partial {#2}}}}

\def\gr{$\gamma$-ray }

\ct  An annihilation fountain at the Galactic center?\par

\ca Martin Pohl\par

\aa DSRI, Juliane Maries Vej 30, 2100 Copenhagen \O, Denmark\par

\ha Abstract\par

\tx Recently, data of taken with the OSSE experiment have been
combined with scanning observations by TGRS and SMM to produce
maps of the narrow Galactic 511 keV line emission (Purcell et al. 1997). 
A modelling of the combined
data give evidence for three distinct features: the Galactic plane, a
central bulge, and an extended emission region at positive latitudes
above the Galactic center.  
It has been proposed that the high-latitude feature is 
associated with a fountain of radioactive debris produced by enhanced
supernova activity in the Galactic center region (Dermer and Skibo 1997).

Here we discuss this scenario in more detail:
we have build a 2-dimensional code to follow the development
of a hydrodynamical fountain in the Galactic center region. We have then
calculated the transport, cooling, and annihilation of positrons as
test particles in the outflow.
As a result we find difficulties with the fountain model if the positrons
are produced by supernovae in a starburst near the Galactic center.

Annihilation on dust grains may increase the 511 keV line flux at high
latitudes. Alternatively the observed positrons may not be entirely produced
by supernovae. 

\ha The model\par

\tx
The basic code is similar to earlier attempts to model
activity in starbursts galaxies (Tomisaka and Ikeuchi
1986, 1988). It is two-dimensional and it assumes azimuthal
symmetry.
The cooling function is based on the work of
Dalgarno and McCray (1972) and Raymond, Cox and Smith (1976).
For the gravitational potential we use the model
of Miyamoto and Nagai (1975), but here with readjusted
parameters (Paczy\'nski 1990).

Direct energy and matter input from supernova explosions is  
described preserving the stochastical nature of these events.
Each supernova is assumed to add $10^{51}\ {\rm ergs}$ of heat
and $10\ M_\odot$ of gas to the interstellar medium.
The number of explosions per time step is determined by a Poissonian
random number generator.
For the general supernova heating, not the starburst, the location of each
supernova is determined by random
number generators for a $r^{-1/2}$ distribution within the Galactic disk and
a normal distribution of dispersion 60 pc in vertical direction.
The hydrodynamical equations are solved with a staggered leapfrog scheme. 

Positron are considered to be a by-product of the explosive event in
the sense that a certain spectrum is injected instantaneously and the particles
convect with the gas flow, cool and eventually annihilate. The positrons
will thus be treated as test particles.

At higher energies the main energy loss processes are inverse Compton 
scattering, synchrotron emission, brems\-strahlung, and adiabatic cooling.
At lower energies the evolution depends on the ionization state of the
background medium. For simplicity we consider gas at temperatures
$T \ge 10^4\ {\rm K}$ fully ionized, below 5000 K the gas is assumed
neutral, and in the intermediate regime 10 \% ionized.
In the former case cooling occurs by Coulomb interactions.
If the background medium is neutral,
the positrons loose energy by ionising and exciting atoms.
A comprehensive summary of the relevant positron annihilation processes
is given by Guessoum, Skibo and Ramaty (1997).

The cooling of the non-relativistic positrons occurs on time scales of
years or less, much shorter than the hydrodynamical time scales.
Therefore we treat the evolution of the positron spectrum semi-analytically
(Drachman 1983).
Having determined the total energy loss rate $A(E)$ and the catastrophic
loss rate $B(E)$, we find that within a time period $\delta t$
(e.g. the timestep of the hydrodynamical calculation) positrons at
an initial energy $E_i$ have cooled to the energy $E_f$, where
$$
\delta t = \int_{E_i}^{E_f} \ {{dE'}\over {A(E')}}\ \eqno(1)
$$
The initial positron spectrum $N_{pos}(E_i)$ has evolved into
$$
N_{pos}(E_f) = N_{pos}(E_i)\ \exp\left(
\int_{E_i}^{E_f} dE'\ {{B(E')}\over {A(E')}}\right)\ \eqno(2)
$$
Since the positrons are test particles in the hydrodynamical flow,
their spatial propagation can be simply described by the number conservation
equation using the grand scale velocity field.

\ha Results\par

\tx We have investigated three basic scenarios:
example 1 considers a moderate general supernova rate of 2 per Millenium
with 1 kpc of the Galactic center. For a short time of half a million
years a starburst occurs at a supernova rate of 20 per Millenium within 500 pc
of the Galactic center. The starburst is centered 60 pc above the midplane,
which in this case is sufficient to establish a one-sided outflow.
The low general supernova rate causes a rather thin gas disk with
little material in the halo. The strong starburst then causes rapid
acceleration of gas in the fountain. As an alternative example 2 is based
on the same initial situation, but for a starburst of half the intensity and
twice the duration compared with example 1. 
The vertical velocity of gas in the fountain then is considerably slower.
Example 3 describes a high general supernova rate of 10 per Millenium
with 1 kpc of the Galactic center. This results in a rather thick and
fluffy gas disk. More half a million years a starburst occurs with 10
supernovae per Millenium within 400 pc of the Galactic center.

We find that though positrons can be efficiently convected to large
heights above the disk, they will in general not produce a very strong annihilation signal there. As an example we show in Figure 1
the vertical distribution of positron and annihilation line flux for
example 1.

\centerline{\psfig{figure=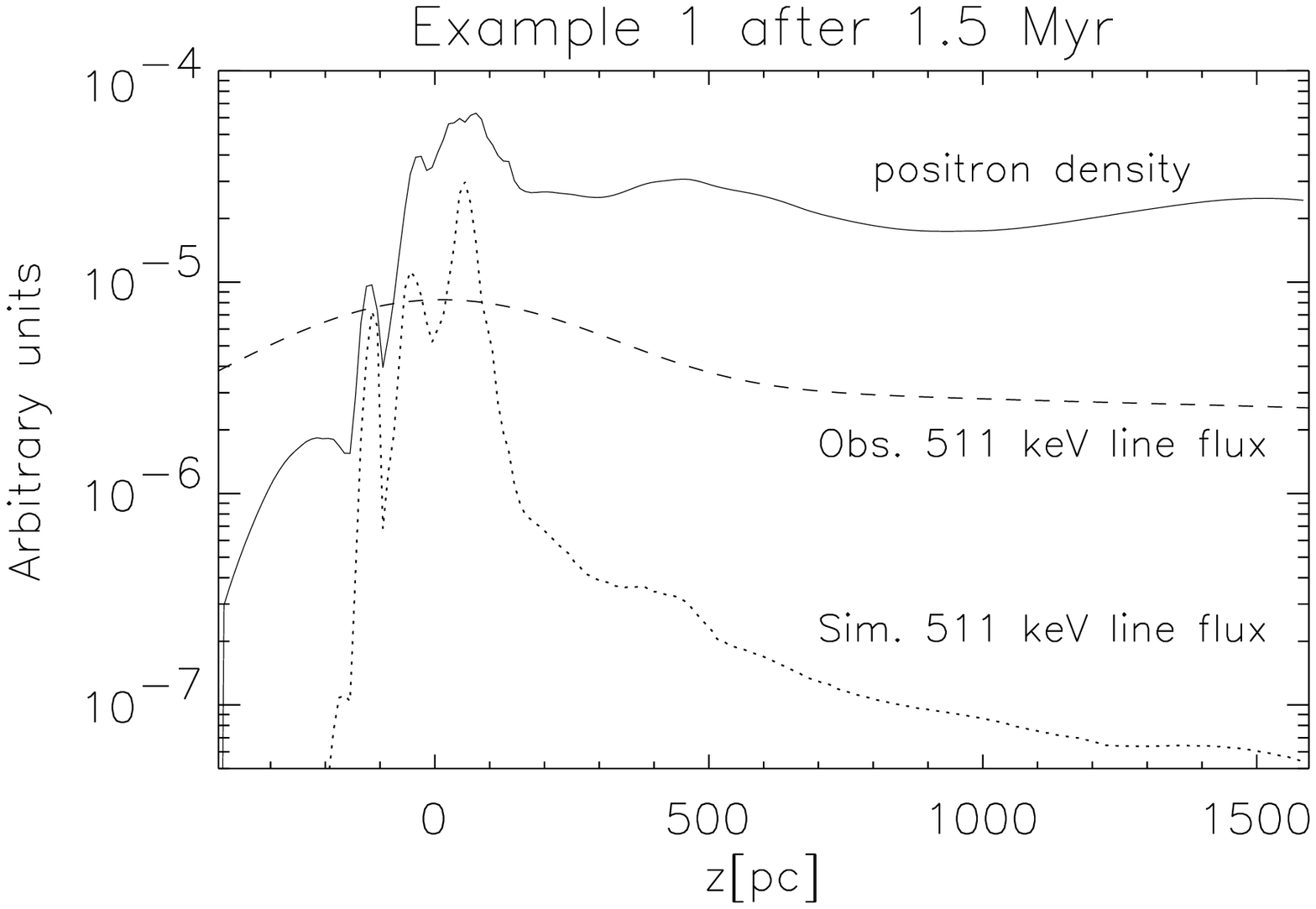,width=14.7cm,clip=}}

\noindent
{\tf Fig.1. The vertical distribution of positron density and narrow 511
keV line flux for example 1, compared with the observed distribution of line flux.
Here we show the status at a simulated time of 1.5
Myr. The starburst occured between 0.1 Myr and 0.6 Myr of
simulated time. Though a fair fraction of positrons has been convected
to large distances above the Galactic plane, the 511 keV line flux is
an order of magnitude less than observed.} 

\vskip15pt

\tx The typical hydrodynamical outflow in our simulations resembles
a narrow cone located exactly above the Galactic center. The initial density
field times the gradient of the gravitational potential $\rho \nabla \Phi$ 
actually confines the flow,
which then rapidly accelerates as it moves outwards. In contrast to the idealized flow envisaged by Dermer and Skibo, the positrons are not injected
in a pre-existing flow, but co-spatial with the pressure input.
Therefore only
a fraction of them will be convected to large heights above the plane. The bulk of positrons annihilates in the disk. 

In our simulations
the typical gas density between 0.5 kpc and 1 kpc above the Galactic plane
is $10^{-2}\ {\rm cm^{-3}}$. The material has temperatures
in excess of $10^6$ K, so that the annihilation rate is
$\sim 10^{-16}\ {\rm sec^{-1}}$. This implies that the life time of
positrons is considerably longer than the duration of a supposed starburst.

As a result the positrons supplied by the starburst will have a strongly
reduced impact on the spatial distribution of annihilation line emission
compared with the naive estimates. In example 1 the ten thousand supernovae
of the starburst result in a high-latitude ($z > 400\ $pc) afterglow of
511 keV line emission with a flux of $3\cdot 10^{-5}\ {\rm cm^{-2}\,sec^{-1}}$,
which is 5\% of the total 511 keV flux 1 Myr after the starburst has ceased.
This is an order of magnitude less than the observed high-latitude flux,
which is about 50\% of the total 511 keV line emission from the Galactic
center region.

We have so far not considered dust at large heights above the Galactic plane.
Zurek (1985) suggested that dust grains may in fact be
quite efficient annihilation sites. Dust expelled by the supernovae and
their progenitor winds may provide annihilation rates considerably higher than
the $\sim 10^{-16}\ {\rm sec^{-1}}$ achieved in our simulations at
locations between 0.5 kpc and 1.5 kpc above the Galactic plane.

\ha Summary\par

\tx In this paper we have investigated in detail the fountain model for the
extraplanar 511 keV line emission observed with OSSE.
If the outflow is caused by starburst activity, the supernovae would provide
both the positron injection and the input of heat and kinetic energy into the
ISM. We have build a two-dimensional code to follow the development of a
hydrodynamical fountain in the Galactic center region. We have then
calculated the transport, cooling, and annihilation of positrons as
test particles in the outflow.

We find difficulties with the fountain model if the positrons
are produced by supernovae in a starburst near the Galactic center.
Even when a strong one-sided outflow of hot gas is established and 
positrons are effectively transported to a distance of 1
kpc above the Galactic plane, the efficiency
of annihilation in this region is very low. 
Almost all of the positrons would annihilate close to the Galactic plane,
and thus they would contribute only little to the high latitude feature of
511 keV line emission.

Annihilation on dust grains may increase the 511 keV line flux at high
latitudes. Alternatively the observed positrons may not be entirely produced
by supernovae. Possible sources of positrons includes black holes or a
\gr burst, which would have to be located above the Galactic center.
It can also not be excluded that the high latitude structure is of
local origin, e.g. from Gould's belt, and not associated with the Galactic 
center. In this case we would expect that high latitude 511 keV line emission
is also observable from other directions.

\ha References\par

\tx
A. Dalgarno, R.A. McCray. {\twelveit A.R.A.\&A.} 10 (1972) 375-426

C.D. Dermer, J.G. Skibo. {\twelveit Ap.J.} 487 (1997) L57-60

R.J. Drachman. {\twelveit Positron-Electron Pairs in 
Astrophysics}. Eds. Burns, Harding and Ramaty. AIP Conf. Proc. 101 (1983) 242-249

N. Guessoum, J.G. Skibo, R. Ramaty. {\twelveit The Transparent Universe}.
Proceedings of the 2nd INTEGRAL workshop. ESA SP-382 (1997) 113-118

M. Leventhal, C.J. MacCallum, P.D. Stang. {\twelveit Ap.J.} 225 (1978) L11-14

M. Miyamoto, R. Nagai. {\twelveit P.A.S.J.} 27 (1975) 533-543

B. Paczy\'nski. {\twelveit Ap.J.} 348 (1990) 485-494

W.R. Purcell et al. {\twelveit Ap.J.} 491 (1997) 725-748

J.C. Raymond, D.P. Cox, B.W. Smith. {\twelveit Ap.J.} 204 (1976) 290-292

K. Tomisaka,S.  Ikeuchi. {\twelveit P.A.S.J.} 38 (1986) 697-715

K. Tomisaka, S. Ikeuchi. {\twelveit Ap.J.} 330 (1988) 695-717

W.H. Zurek. {\twelveit Ap.J.} 289 (1985) 603

\vfill\eject
\end\bye